# Quantum coherence control at near 1000 K

Gang-Qin Liu[1,*], Xi Feng[1,*], Ning Wang[1], Quan Li[1,2,3,†], Ren-Bao Liu[1,2,3,†]

[1] Department of Physics, The Chinese University of Hong Kong, Shatin, New Territories, Hong Kong, China

[2] The Chinese University of Hong Kong Shenzhen Research Institute, Shen Zhen 518100, China

[3] Centre for Quantum Coherence, The Chinese University of Hong Kong, Shatin, New Territories, Hong Kong, China

* These authors contributed equally

† Correspondence should be addressed to R.B.L. (rbliu@cuhk.edu.hk) or Q.L.(liquan@phy.cuhk.edu.hk)

**Abstract:** Quantum coherence control usually requires extremely low temperature environments. Even for spins in diamond, a remarkable exception, the coherence signal is lost as temperature approaches 700 K. Here we demonstrate quantum coherence control of the electron spins of nitrogen-vacancy centers in nanodiamonds at temperatures near 1000 K. The scheme is based on initialization and readout of the spins at room temperature and control at high temperature, which is enabled by pulse laser heating and rapid diffusion cooling of nanodiamonds on amorphous carbon films. Using high-temperature spin control, we observe the magnetic phase transition of a single nickel nanoparticle at about 615 K. This work enables nano-thermometry and nano-magnetometry in the high-temperature regime.

**Main text:**

Quantum phenomena usually occurs only at low temperature. Quantum coherence control at high temperature would enable quantum sensing[1] in many important fields of nanoscience and nanotechnology, such as thermoremanent magnetization of nanoparticles[2], magnetic records in petrology and planetary science[3], thermoelectric effects in nanostructures[4–7], heat-assisted magnetic recording[8,9], and thermo-plasmonics of nanoparticles[10–12], where the relevant temperature is often > 500 K, with few nano-sensors available. Remarkably, the nitrogen-vacancy (NV) center in diamond has been demonstrated to have robust quantum coherence at room temperature[13,14] and even up to 700 K[15], which has stimulated enormous studies for quantum information processing[14] and quantum sensing[1,16]. However, at temperature > 700 K, even the NV center loses its coherence signals[15].

Here we demonstrate quantum coherence control of the electron spins of NV centers in nanodiamonds at temperatures near 1000 K. The scheme is based on initialization and readout at room temperature and control at high temperature, which is enabled by pulse laser heating and rapid diffusion cooling of nanodiamonds on amorphous carbon films. Using a nanodiamond sensor working at high temperature, we observe the magnetic phase transition of a single nickel nanoparticle at about 615 K. This work, integrating high-temperature quantum control and nanoscale resolution, offers an approach to nano-thermometry and nano-magnetometry in the high-temperature[2–12].

We measure optical detection of magnetic resonance [13] (ODMR) of NDs deposited on the amorphous carbon films on TEM copper grids (Fig. 1A). The ground state of NV centers is a spin-triplet with the spin quantum number along the NV axis $m_S = 0$ or $\pm 1$, with a zero-field splitting (ZFS) between $|0\rangle$ and $|\pm 1\rangle$ being $D \approx 2.87$ GHz at room temperature (Fig. 1B). The ZFS depends on temperature $T$ with $dD/dT = -74$ kHz/K near room temperature [17], which can be used to calibrate the temperature. The NV center can be excited by a laser to a triplet state with $m_S$ conserved and then return to the ground state either by spin-conserving photon emission or by non-radiative relaxation via inter-system crossover to the intermediate singlet states (Fig. 1B). The non-radiative process is more efficient if the NV is in the $|\pm 1\rangle$ states and the intermediate singlet states have higher probability to relax to the $|0\rangle$ state, so a laser pulse can initialize NV centers to the $|0\rangle$ state in a few microseconds and the $|0\rangle$ has stronger photoluminescence than the $|\pm 1\rangle$ states. Between the initialization and photon-counting readout, microwave pulses can be applied to coherently manipulate the spin state. Therefore, the photon counts depend on the frequency of the microwave pulses, hence the ODMR. In previous experiments, it was found that both the overall photoluminescence intensity and the contrast between different spin states drop dramatically as the temperature increases above 550 K (see Methods & Supplementary Information), making ODMR of NV centers invisible at above 700 K [15]. The loss of spin-resonance signal at high temperature was ascribed to the enhancement of non-radiative relaxation of optically excited NV center state with $m_S = 0$, which prevented the spin readout at high temperature [15]. To overcome the temperature limit, we initialize and read out the spins at low temperature (< 500 K) and manipulate them at high temperature.

The scheme is shown in Fig. 1C. The carbon films absorb light and can be rapidly heated in a local area by a focused laser pulse and in turn heat the ND near the focus spot (see color map of Fig. 1A). Since the heating spot is small, when the heating laser is switched off the ND will rapidly cool down by heat diffusion through the carbon film. We use a weak 532 nm green laser pulse to initialize the NV centers to the $|0\rangle$ state, and then apply a strong 808 nm near-infrared (NIR) laser pulse for rapid heating, and finally carry out the readout by another weak 532-nm laser pulse after the ND has cooled down. A short (<100 ns) microwave pulse is applied between the initialization and readout to manipulate the spin state. The 808-nm wavelength for the heating laser is chosen to minimize disturbance of the spin states and the charge states of the NV centers. The peak temperature can be tuned by adjusting the laser power and the pulse duration. To suppress the oxidation of the carbon film (which reduces the heating efficiency) during the laser heating, the sample (after TEM imaging) is sealed in an argon (Ar) gas chamber (see Methods & Supplementary Information).

Figure 1D shows the ODMR spectra of an ND for various heating laser powers. The measurement is under zero magnetic field. The microwave pulse with duration of 30 ns is applied at the end of the 10 μs NIR heating pulse. The ZFS $D$ shifts to lower frequencies with increasing the heating power. The contrast of ODMR persists until $D$ is $2758 \pm 1$ MHz. To determine the temperature for different heating laser powers, we use the $D(T)$ curve measured in Ref. [15] for $T <$ 700 K. For higher temperature, we extract the temperature by extrapolation of the exponential cooling after the heating pulse (see Methods & Supplementary Information). As shown in the lower panes of Fig. 1C, the heating and cooling dynamics, obtained by varying the heating pulse duration ($\tau_h$) and the waiting time ($t_w$) of the microwave pulse, are both well fitted by a rate equation with a heating rate (which depends on the laser power) and a cooling rate (due to heat diffusion through the carbon film) (see Methods & Supplementary Information). The temperature immediately after the heating pulse is determined by extrapolation of the exponential cooling using the temperatures at different $t_w$ (which are calibrated by $D(T)$ as they are below 700 K). Figure 1E shows the temperature dependence of the ZFS $D(T)$ obtained by the extrapolation approach, in comparison with the result from Ref. [15]. With temperature thus calibrated, the ODMR is observed at temperature as high as $1004 \pm 24$ K (Fig. 1D).

We now measure the spin relaxation time of the NV centers at high temperature. We initialize the spins in the $m_S = -1$ or $m_S = 0$ state by a 532-nm laser pulse with or without a microwave $\pi$ pulse right after the laser pulse (see Methods & Supplementary Information). Then an NIR heating pulse is applied and the spin population is measured after a fixed cooling time $t_c$. Figure 2A plots the difference between the photon counts for the $m_S = -1$ and $m_S = 0$ initial states as a function of the heating duration $\tau_h$ for various heating powers, showing the spin relaxation for various temperatures. At high temperature, the spin relaxation time $T_1$ (Fig. 2B) is well fitted by $T_1^{-1} = AT^n$ with $A = 1.9 \times 10^{-12}$ s$^{-1}$K$^{-n}$ and $n = 5.83 \pm 0.6$, which is close to the temperature dependence due to the two-phonon Raman processes ($\propto T^5$) [18]. The spin relaxation time saturates at ~100 microseconds at room temperature, which is ascribed to the electric and magnetic noises in the NDs (Supplementary Information).

We demonstrate quantum control of the NV center spins at high temperature by Rabi oscillation. A weak magnetic field (38 Gauss) is applied to split the resonances of the NV centers along the four different crystallographic orientations (Fig. 2C). The Rabi oscillation is observed by applying a microwave pulse resonant with the first transition (with the lowest frequency) in Fig. 2C (which has the largest contrast). Though its visibility decreases with increasing the temperature due to the spin relaxation during the heating process, the Rabi oscillation persists at temperature up to 938 K with no significant change of the coherence time (Fig. 2D).

A potential application of high-temperature ODMR of NDs is to study the magnetism of single nanoparticles, where the Curie temperatures or superparamagnetic blocking temperatures are often above 500 K, beyond the working temperature range of most existing nano-sensors. We demonstrate the high-temperature nano-magnetometry by measuring the magnetic properties of a single nickel nanoparticle (Ni NP) [19]. As shown in the inset of Fig. 3B, an ND located near a Ni NP is used as the sensor. The stray magnetic field from the Ni NP causes splitting and broadening of the NV center spin resonances (Fig. 3A). The broadening is due to the gradient of the stray field in the range of the ND which contains about 500 randomly distributed NV centers [19]. With increasing the temperature, the ODMR spectra become narrowed, indicating the demagnetization of the Ni NP. The ODMR splitting (Fig. 4B) due to the Ni NP is measured in four rounds of cooling processes from a high temperature $T_H$ to a low one $T_L$. A magnetic phase transition at $T_C = 615 \pm 4$ K is clearly seen. When the Ni NP is cooled down from above to below $T_C$ (round 1 & 3 in Fig. 3B, in which $T_H > T_C > T_L$), the resultant magnetization at low temperature is random due to the spontaneous symmetry breaking in the phase transition.

The high-temperature ODMR opens new opportunities of nano-thermometry and nano-magnetometry. Using the photon counts (8 Mps), the ODMR width (10 MHz, mainly due to ensemble broadening), and contrast (0.05) of the NDs we have measured at 1000 K, we estimate the temperature sensitivity to be about 250 mK/Hz$^{1/2}$ based on the ZFS shift $dD/dT \approx 240$ kHz/K (determined from data in Fig. 1E) and the magnetic field sensitivity to be 2.5 µT/Hz$^{1/2}$ (see Methods & Supplementary Information). The temperature range of the spin coherence control in diamond can be pushed to even higher. In our current experiment, the spin relaxation time (~ 5 µs) at about 1000 K is comparable to the laser heating time (~ 2 µs) in our setup, making it challenging to observe the ODMR above 1000 K. This limit can be overcome by either increasing the heating and cooling speed or by increasing the spin relaxation time. The latter can be realized, e.g., by transferring the polarization of electron spins to that of nuclear spins [20].

## METHODS

**Samples** – NDs with ensemble NV centers (average size ~ 140 nm) are purchased from Adámas. Each ND contains about 500 NV centers. NDs dispersed on a TEM grid (Ted Pella) are attached to the amorphous carbon film. NDs are stable on the carbon film, with no change observed during the whole measurement. Ni NPs are obtained by ball milling of Ni powder (3-7 um, Strem Chemicals). High temperature annealing (973 K, 2 hours) is performed in 10% H2/Ar after ball milling to improve the crystallinity of the NPs. After size sorting by centrifugation, Ni NPs with size of about 100 nm are dispersed on a TEM grid. The composition of single Ni NPs is verified by energy-dispersive X-ray spectroscopy (EDX). The locations of Ni NPs and NDs in the proximity are identified by TEM imaging using the unique net patterns of the amorphous carbon film. In confocal fluorescence imaging, the NDs are much brighter than the carbon film and the Ni NPs. The overlap between the TEM and confocal images is employed to locate NDs with Ni NPs in the proximity.

**Sample holder** – To avoid oxidation of the amorphous carbon film during laser heating, the sample is protected in an argon (Ar) atmosphere in the high-temperature measurements. The chamber is built on a confocal dish, with a reusable cover. The bottom of the dish is removed, then glued to the PCB board with microwave transmission lines. The TEM grid is fixed in the chamber, and then the opened chamber is placed in glovebox of an Ar atmosphere for about 10 hours to be filled with Ar. Finally, the chamber is closed with the dish cover and sealed with additional glue before being taken out from the glovebox for ODMR measurement.

**ODMR setup** – The optical system contains (1) spin polarization with a green laser (532 nm), (2) spin readout with fluorescence collection under 532 nm laser excitation, and (3) local and instantaneous heating with an NIR laser (808 nm). The 532-nm laser is modulated by an acousto-optic modulator (AOM) and coupled into a microscopy frame through a single-mode fiber. A pair of galvo mirrors control the focusing position (X-Y) of the green laser, and a piezo stage controls the focus depth of the oil objective (NA=1.35). The fluorescence of NV centers is collected by the same objective and passed through two dichroic mirrors (one for the green laser and another for the NIR) and filters, then detected by a single photon counting module (SPCM, Excelitas). The NIR laser is independently controlled with an AOM and a pair of galvo mirrors, adjusted to overlap with the green laser. A pair of moving lenses are used to compensate the chromatic aberration between the two lasers. The amplitude and frequency of MW pulses are controlled by the signal generator (N5181A, Agilent) and their shapes are modulated by a RF switch (ZASW-2-50DR, Mini Circuits). The laser excitation, NIR heating, MW manipulation, and fluorescence readout are synchronized with TTL signals from a pulse generator (PulseBlasterESR-PRO, SpinCore).

**Temperature calibration** – The temperature below 700 K is calibrated with the *D-T* relation from Ref. [15]. For temperature higher than 700 K, we measure the ZFS *D* at different waiting times $t_w$ (see Fig. 1C) and then use extrapolation to determine the temperature right after the heating pulse with the assumption of exponential cooling (which is verified in experiments). In Rabi oscillation experiments, the *D* value is determined by averaging the frequencies of the two outmost resonances and then the *D-T* curve in Fig. 1E is used to determine the temperature.

**More details are available in the Supplementary Information.**

**REFERENCES**

1. Degen, C. L., Reinhard, F. & Cappellaro, P. Quantum sensing. *Rev. Mod. Phys.* **89,** (2017).
2. Dunlop, D. J. & Özdemir, Ö. Rock Magnetism Fundamentals and Frontiers. *Cambridge Univ. Pr* (1997).
3. Fu, R. R. *et al.* Solar nebula magnetic fields recorded in the Semarkona meteorite. *Science (80-. ).* **346,** 1089–1092 (2014).
4. Poudel, B. *et al.* High-thermoelectric performance of nanostructured bismuth antimony telluride bulk alloys. *Science (80-. ).* **320,** 634–638 (2008).
5. Zhao, W. *et al.* Magnetoelectric interaction and transport behaviours in magnetic nanocomposite thermoelectric materials. *Nat. Nanotechnol.* **12,** 55–60 (2017).
6. Zhao, W. *et al.* Superparamagnetic enhancement of thermoelectric performance. *Nature* **549,** 247–251 (2017).
7. Zhou, X. *et al.* Routes for high-performance thermoelectric materials. *Materials Today* (2018).
8. Challener, W. A. *et al.* Heat-assisted magnetic recording by a near-field transducer with efficient optical energy transfer. *Nat. Photonics* **3,** 220–224 (2009).
9. Kryder, M. H. *et al.* Heat Assisted Magnetic Recording. *Proc. IEEE* **96,** 1810–1835 (2008).
10. Baffou, G. *Thermoplasmonics*. (Cambridge University Press, 2017).
11. Boyer, D., Tamarat, P., Maali, A., Lounis, B. & Orrit, M. Photothermal imaging of nanometer-sized metal particles among scatterers. *Science (80-. ).* **297,** 1160–1163 (2002).
12. Brongersma, M. L., Halas, N. J. & Nordlander, P. Plasmon-induced hot carrier science and technology. *Nat. Nanotechnol.* **10,** 25–34 (2015).
13. Gruber, A. *et al.* Scanning Confocal Optical Microscopy and Magnetic Resonance on Single Defect Centers. *Science (80-. ).* **276,** 2012–2014 (1997).
14. Doherty, M. W. *et al.* The nitrogen-vacancy colour centre in diamond. *Phys. Rep.* **528,** 1–45 (2013).
15. Toyli, D. M. *et al.* Measurement and control of single nitrogen-vacancy center spins above 600 K. *Phys. Rev. X* **2,** (2012).
16. Rondin, L. *et al.* Magnetometry with nitrogen-vacancy defects in diamond. *Reports on Progress in Physics* **77,** (2014).
17. Acosta, V. M. *et al.* Temperature Dependence of the Nitrogen-Vacancy Magnetic Resonance

in Diamond. *Phys. Rev. Lett.* **104,** 070801 (2010).
18. Norambuena, A. *et al.* Spin-lattice relaxation of individual solid-state spins. *Phys. Rev. B* **97,** 094304 (2018).
19. Wang, N. *et al.* Magnetic Criticality Enhanced Hybrid Nanodiamond Thermometer under Ambient Conditions. *Phys. Rev. X* **8,** 011042 (2018).
20. Dutt, M. V. G. *et al.* Quantum register based on individual electronic and nuclear spin qubits in diamond. *Science (80-. ).* **316,** 1312–1316 (2007).

**Acknowledgments -** This work was supported by the National Basic Research Program of China (973 Program) under Grant No. 2014CB921402, Hong Kong Research Grants Council-Collaborative Research Fund Project C4006-17G, Hong Kong Research Grants Council – General Research Fund Project 14319016, and The Chinese University of Hong Kong Group Research Scheme 2017-2018 Project 3110126.

**Author contributions -** R.B.L. conceived the idea. R.B.L. & Q.L. supervised the project. R.B.L., Q.L. & G.-Q. L. designed the experiment. G.Q.L. noticed the laser heating effect of carbon thin films. G.-Q. L. & N.W. set up the HiT-ODMR system and carried out the ODMR experiments. G.Q.L. prepared the NDs & the Ar-filled sample chambers. X.F. characterized NDs on carbon thin films, and prepared and characterized Ni NPs, and NDs & Ni NPs on carbon films. G.-Q. L., R.B.L. & Q.L. analyzed the data. G.-Q. L., R.B.L. & Q.L. wrote the paper. All authors commented on the manuscript.

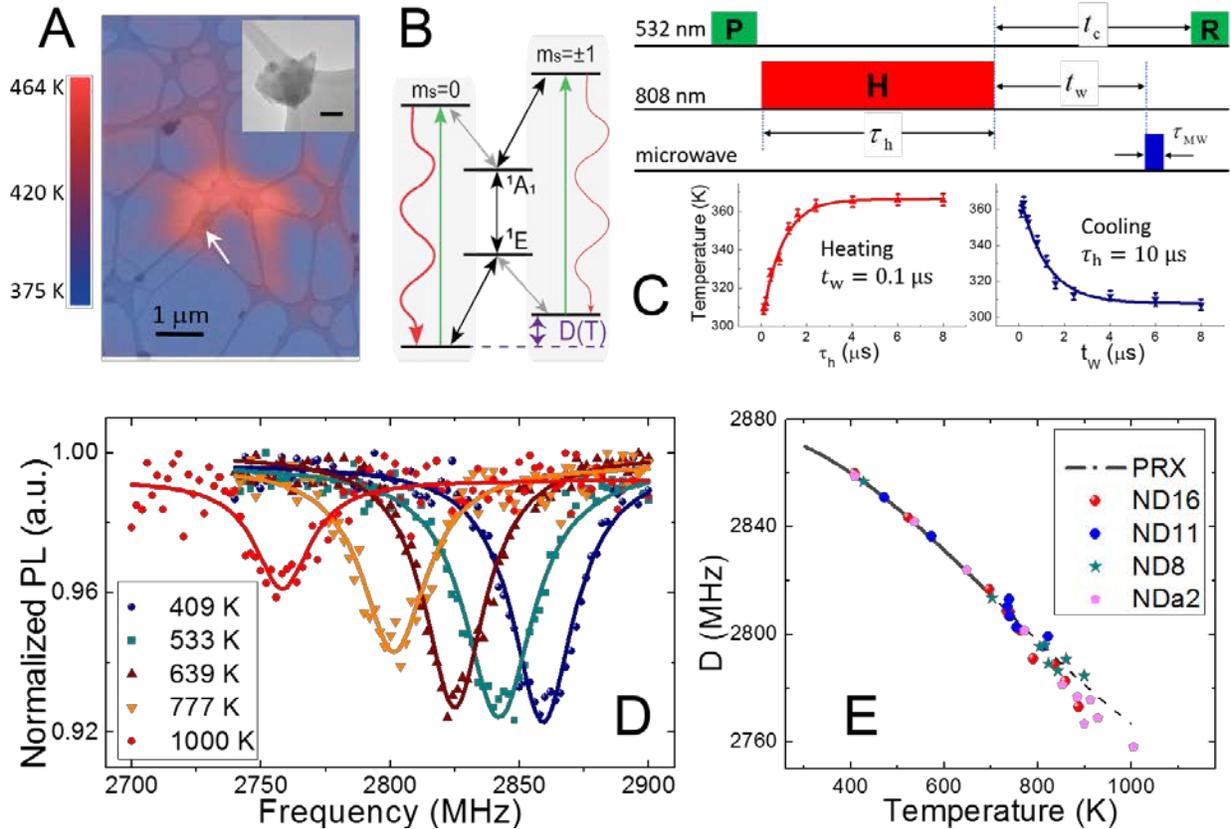

**Figure 1 | ODMR of NV centers in nanodiamonds at temperature approaching 1000 K.** (A) TEM image of amorphous carbon film with NDs and Ni NPs. The background color shows the temperature of an ND (indicated by the arrow) as a function of the position of the NIR laser focus spot. Inset: A typical TEM image of an ND on a carbon film (scale bar: 50 nm). (B) The energy level diagram of an NV center illustrating the spin polarization and readout. (C) Timing of the 532 nm laser pulses (for polarization and readout, P/R), the 808 nm laser pulse for heating (H), and the microwave pulse for spin control. The lower panes show the heating (left) and cooling (right) dynamics by plotting the temperature of the ND as a function of the heating pulse duration $\tau_h$ with the waiting time fixed to be $t_w = 0.1$ μs or as a function of the waiting time $t_w$ for a fixed $\tau_h = 10$ μs. (D) ODMR spectra of an ND for various heating laser powers (corresponding temperatures indicated). The heating pulse duration is fixed to be $\tau_h = 5$ μs and the waiting time $t_w = 0$ μs. (E) The temperature dependence of the ZFS $D(T)$ obtained by extrapolation of the exponential cooling dynamics for four different NDs (indicated by different symbols), in comparison with the results from Ref. [15] (solid line for $T < 700$ K). The dashed line extends the fitting formula in Ref. [15] to 1000 K.

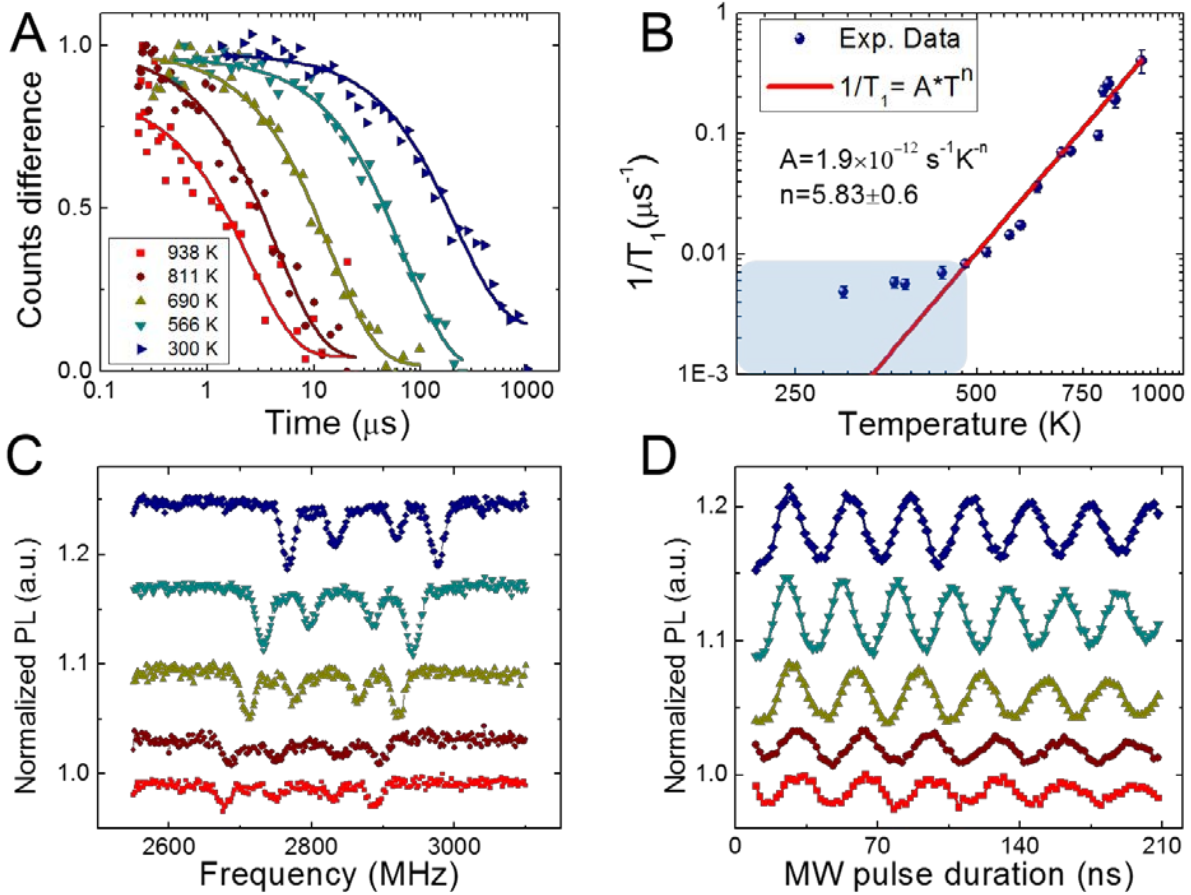

**Figure 2 | Quantum coherence of NV center spins in nanodiamonds at temperatures approaching to 1000 K.** (A) Relaxation of ensemble NV center spins of an ND (NDa2 in Fig. 1E) measured at different temperatures. (B) Spin relaxation rate ($1/T_1$) as a function of temperature. The red line is the power-law fitting in the high temperature regime. At lower temperature (shadowed region), the relaxation rate saturates. (C) ODMR spectra and (D) Rabi oscillations of ensemble NV center spins in the ND at different temperatures (indicated by different symbols as shown in A). An external magnetic field of about 38 Gauss is applied in those measurements. The temperature is determined by the ZFS from the $D(T)$ relation in Fig. 1E.

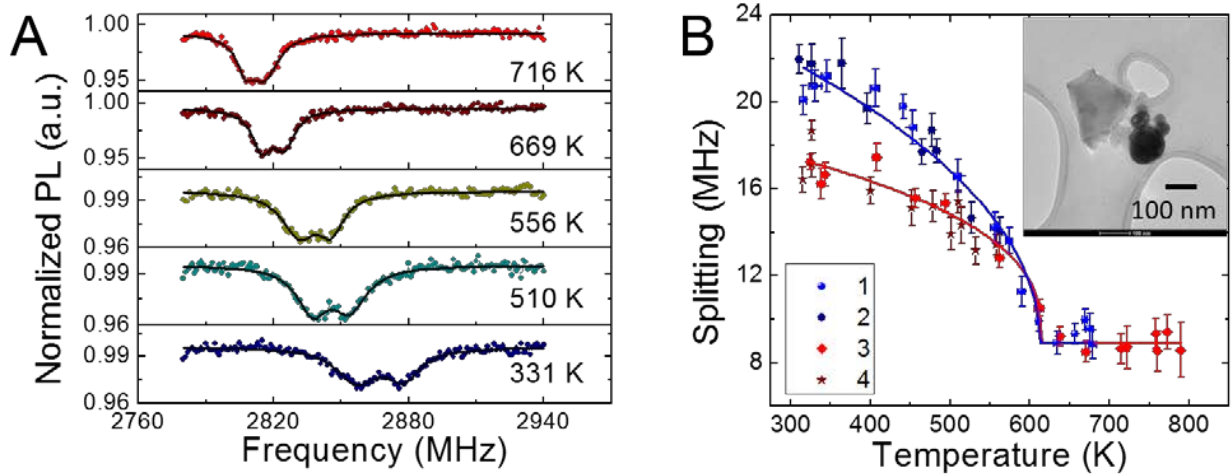

**Figure 3 | Magnetic phase transition of a single nickel nanoparticle.** (A) Typical zero-field ODMR spectra of an ND located close to a Ni NP. The splitting and extra broadening are induced by the magnetization of the nearby Ni NP. (B) Spontaneous magnetization of the Ni NP in several rounds of cooling, the measurement order is Round 1 ($T_H$ = 669 K to $T_L$ = 316 K) → Round 2 (563 K to 311 K) → Round 3 (789 K to 324 K) → Round 4 (558 K to 326 K). The magnetization in a cooling process is nearly the same as the previous round if the cooling starts at a relative low temperature ($T_H < T_C$, Round 2 and 4), or otherwise different from the previous round ($T_H > T_C$, Round 3). The solid lines are fitting curves with a function of $\sim (1 - T/T_C)^\beta$ to extract $T_C$ (615 ± 4 K) of this Ni NP. Inset: TEM image of the measured ND (light gray) and the nearby Ni NP (black sphere).

# Supplementary Information

## 1. Nanodiamond and Ni nanoparticle.

Nanodiamonds (NDs) with ensemble NV centers were purchased from Adámas, with initial concentration of 1 mg/ml (water solution). The average size of NDs is about 140 nm from dynamic light scattering (DLS) measurement. Each ND contains about 500 NV centers. To prepare the sample for high-temperature ODMR (HiT-ODMR), a drop of 10 μl ND ethanol solution (5 μg/ml) was transferred to a TEM grid (Ted Pella) with pipette. NDs attached to the amorphous carbon film by Van der Waals' force after the ethanol volatilized (in several minutes, see Fig. 1A inset of the main text for a typical TEM image of a bare ND on the carbon film). ND appears to be stable on the carbon film, with no change observed during the whole measurement.

Nickle nanoparticles (NPs) were obtained by ball milling of Ni powder (3-7 um, Strem Chemicals). High temperature annealing (973 K, 2 hours) was performed in 10% $H_2$/Ar after ball milling to improve the crystallinity of the NPs. After size sorting by centrifugation, Ni NPs with size of about 100 nm were dispersed in ethanol and dropped on a TEM grid with the same method as for depositing NDs on a TEM grid. Figure S1 shows a typical TEM image of a single Ni NP and its energy-dispersive X-ray spectroscopy (EDX) data.

The locations of Ni NPs and NDs in the proximity were identified by TEM imaging, as shown in Fig. S2 below and Fig. 3B of the main text. On a TEM grid, the amorphous carbon film forms unique net patterns (Fig. S2A), which serves as the natural markers to locate the nanoparticles in the confocal image. The confocal image (Fig. S2B) shows that the NDs are much brighter than the carbon film and the Ni NPs. The overlap between the TEM and confocal images (Fig. S2C) was employed to locate NDs with Ni NPs in the proximity (Fig. S2D).

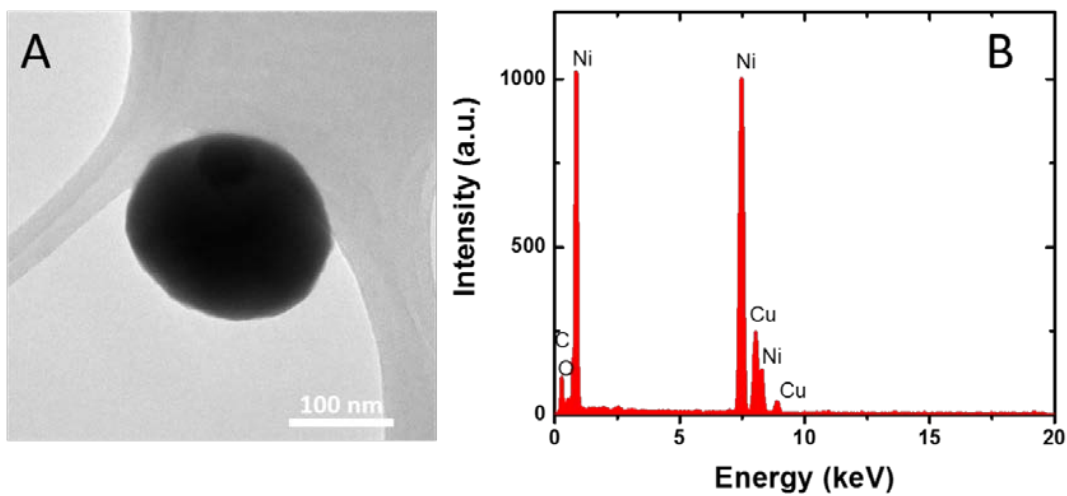

**Figure S1 | A typical Ni nanoparticle on a carbon film.** (**A**) The TEM image of a Ni NP (black sphere) on a carbon film (gray frame). (**B**) The EDX spectrum shows that the NP is pure Ni, with Cu and carbon signals coming from the TEM grid.

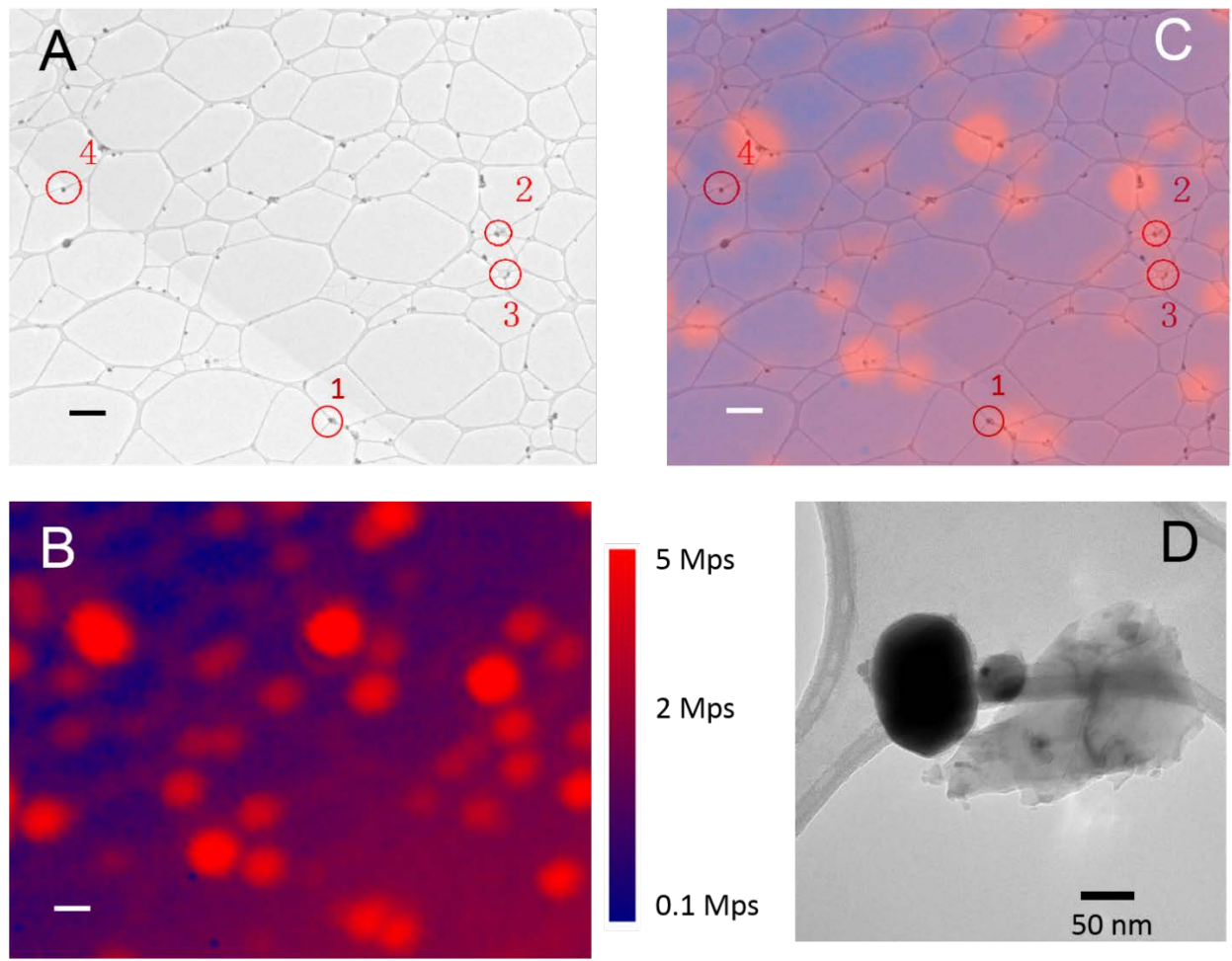

**Figure S2 | Identifications of NDs and Ni NPs by overlap of TEM and confocal images.** (**A**) The TEM image of the NDs and Ni NPs on a carbon film. The gray mesh pattern is the amorphous carbon film. The darker dots on the carbon film are ND or Ni NPs (**B**) The confocal image of NDs and Ni NPs on a TEM grid (the same region as in A). Color bar shows the fluorescence counts of each pixel. The bright spots are NDs with ensemble NV centers. The carbon film and the Ni NPs have much weaker fluorescence than the NDs. (**C**) Overlap of the TEM and confocal images. In A, B and C, four Ni NPs with nearby NDs are circled and labelled. The scale bars in A, B and C are 1 μm. (**D**) Close-up TEM image of an ND and Ni NPs (No. 3 in **A**). The Ni NPs (sphere-shaped) have darker contrast than the ND (triangle-shaped).

## 2. Sample chamber

To avoid oxidation of the amorphous carbon film during laser heating, the sample was protected in an argon (Ar) atmosphere in the high-temperature measurements. The chamber was built on a confocal dish, with a reusable cover, as shown in Fig. S3. The bottom of the dish was removed, then it was glued to the PCB board with microwave transmission lines. A 25-μm-diameter copper wire soldered to the transition line buried in between the dish and the PCB board, which was employed to delivery microwave to the sample. The bottom side of the dish was finally closed with a cover glass, which formed the optical window for microscopy. The TEM grid was fixed in the chamber, and then the opened chamber was placed in glovebox of an Ar atmosphere for about 10 hours to be filled with Ar. Finally, the chamber was closed with the dish cover and sealed with additional glue before being taken out from the glovebox for ODMR measurement.

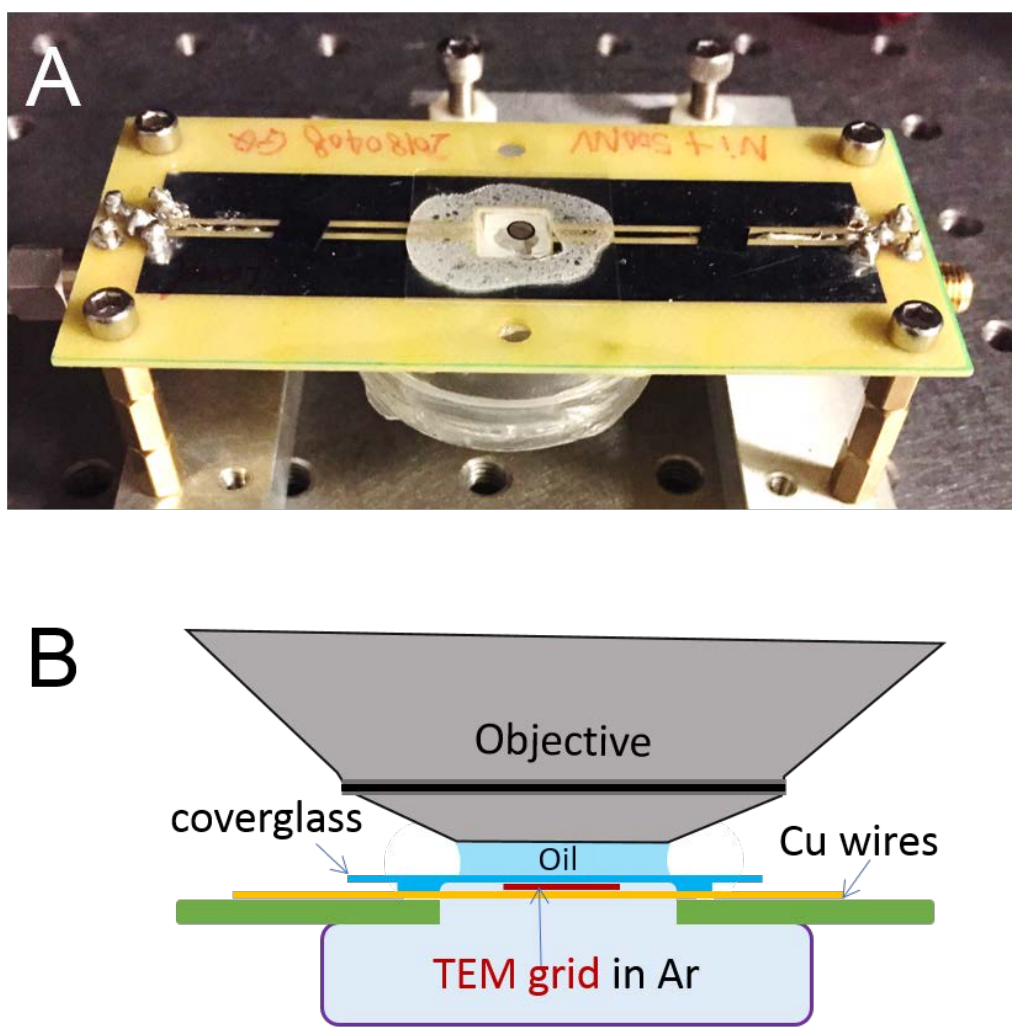

**Figure S3 | A home-built sample chamber mounted on a PCB board.** (**A**) Photograph of the sample chamber. (**B**) Diagram of the sample chamber in ODMR measurement.

## 3. ODMR setup with NIR laser heating

As shown in Fig. S4A, the optical system contained 3 parts: (1) spin polarization with a green laser (532 nm), (2) spin readout with fluorescence collection under 532 nm laser excitation, and (3) local and instantaneous heating with an NIR laser (808 nm). The 532-nm laser was modulated by an acousto-optic modulator (AOM) and coupled into a microscopy frame through a single-mode fiber. A pair of galvo mirrors were used to control the focusing position (X-Y) of the green laser, and a piezo stage was used to control the focus depth (Z) of the oil objective (NA=1.35). The fluorescence of NV centers in NDs was collected by the same objective and passed through two dichroic mirrors (DM, one for the green laser and another for the NIR laser) and filters. The fluorescent signals were converted into digital signals by a single photon counting module (SPCM, Excelitas) and then recorded by a digital counter (USB-6211, National Instruments). The NIR laser was independently controlled with an AOM and a pair of galvo mirrors, adjusted to overlap with the green laser at the DM2 position. A pair of moving lenses (orange dash box) were used to compensate the chromatic aberration between the two lasers.

Microwave (MW) pulses synchronized with the optical pulses were used to manipulate the spin states of the NV centers. The amplitude and frequency of the MW signal were controlled by the signal generator (N5181A, Agilent), and the shape of the MW pulses was modulated by a RF switch (ZASW-2-50DR, Mini Circuits). After amplification, the MW pulses were delivered to the sample through a coaxial cable, transmission lines on the sample holder, and a 25-μm-diameter copper wire. The laser excitation, NIR heating, MW manipulation, and fluorescence readout were synchronized with TTL signals from a pulse generator (PulseBlasterESR-PRO, SpinCore).

Typical zero-field continuous-wave (CW) ODMR spectra of a bare ND with ensemble NV centers are presented in Fig. S4B. The splitting of the peak is caused by the local strain of the diamond lattice. Local temperature was tuned by applying an NIR laser of different power. The shift of the resonant frequencies was induced by NIR laser heating. The temperature depends approximately linearly on the NIR laser power in the measured range (Fig. S4C). When temperature of the ND was close to 550 K ($D < 2840$ MHz), the contrast of the ODMR spectrum begun to decrease (Fig. S4C), similar to observation in Ref. (*1*).

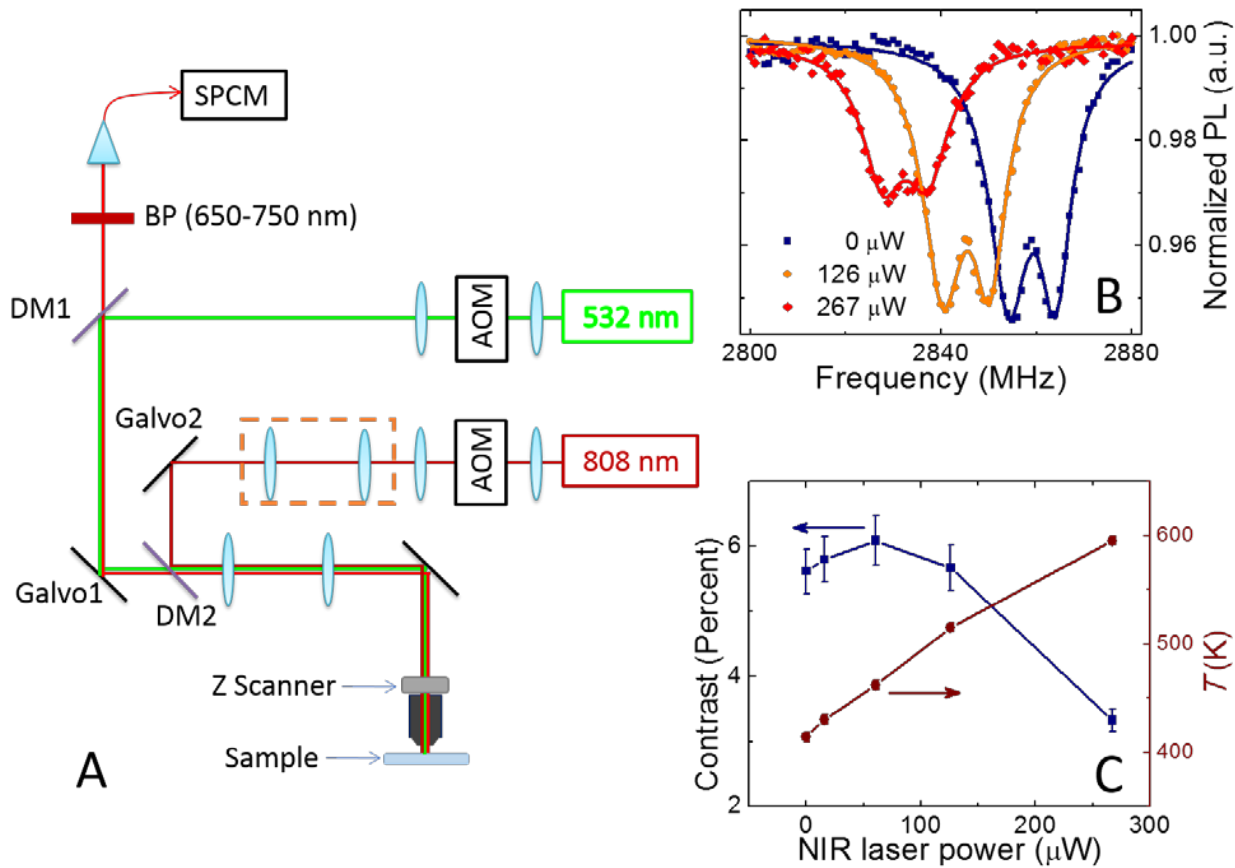

**Figure S4 | ODMR setup and typical ODMR spectra**. (**A**) Diagram of the confocal system. AOM: acousto-optic modulator. DM: dichroic mirrors. BP: bandpass filter. Galvo: Scanning Galvo Mirrors. SPCM: single photon counting module. (**B**) Typical zero-field ODMR spectra of ensemble NV centers in an ND with NIR laser heating of different power. (**C**) The temperature and contrast of ODMR spectra as functions of the NIR laser power.

## 4. NIR laser heating

The NIR laser heating was first characterized by heating imaging. As shown in Fig. 1A of the main text and Fig. S5A, the three-point ODMR (*2*) was measured as the NIR laser beam was scanned around a selected ND. The three microwave frequencies were chosen as: $f_1$ and $f_2$ at the half-maximum points of the spin resonance with the NIR laser turned off and $f_3$ far away from the resonance frequency. The photon counts for the three microwave frequencies $f_{1/2/3}$ are denoted as $c_{1/2/3}$. The relative contrast change $C = (c_1 - c_2)/(2c_3 - c_1 - c_2)$ is an approximatively linear function of the resonance frequency shift if it is not too large (e.g., within in the ODMR width). Thus, the $C$ value was converted to temperature, with calibration by two full ODMR spectra: one measured with the NIR laser turned off and one measured at the highest temperature. The patterns in the heat mapping image (Fig. 1A of main text and Fig. 5SA) correspond to the structures of the amorphous carbon films. With this method, we optimized the position of the NIR laser beam for heating efficiency. Usually the optimal position overlapped with the selected ND (e.g., the black box in Fig. S5A).

We studied the NIR heating dynamics by heat conduction measurement. Figure S5B shows the temperature of an ND (measured with the same sequence as in Fig. 1C of the main text) as a function of the duration of the heating NIR pulse for various NIR laser focus spots on the amorphous carbon film. For each focus spot and each heating duration, a full ODMR spectrum was measured and the zero-field splitting (ZFS) $D$ was extracted by Lorentzian fitting. When the NIR laser was away from the ND position ($P_0$), it took a delay time after the heating pulse for the temperature of the ND to rise, and the stationary temperature was lower (i.e., the final state $D$ was larger). The delay time, which increases linearly with the distance (Fig. S5C), is attributed to the propagation of heat from the heating spot to the ND through the amorphous carbon film. Note that there was an extra delay in the experimental data (shadowed regions of Fig. S5B-C), which was induced by the NIR AOM delay of our setup.

All the $D$ data are well fit with an exponential decay with the decay time nearly the same in the measured temperature range (300 – 380 K). In this temperature range, the ZFS $D$ is an approximately linear function of the temperature. The exponential decay can be well understood with a rate equation. We define the following parameters:

$T_0$ - initial temperature of the ND

$T_E$ - temperature of the environment

$W$ - heating rate, determined by NIR laser power

$\gamma$ - cooling rate, determined by the thermal conductivity of the amorphous carbon film.

The heating/cooling dynamics is determined by the rate equation

$$\frac{dT}{dt} = W - \gamma(T - T_E). \tag{S1}$$

Thus, the temperature of the ND is

$$T = T_E + \frac{W}{\gamma} + \left(T_0 - T_E - \frac{W}{\gamma}\right)\exp(-\gamma t). \tag{S2}$$

The local temperature increases (decreases) exponentially as the NIR laser is turned on (off), with a time scale $\gamma^{-1}$, before it reaches the stationary value ($T_E + W/\gamma$).

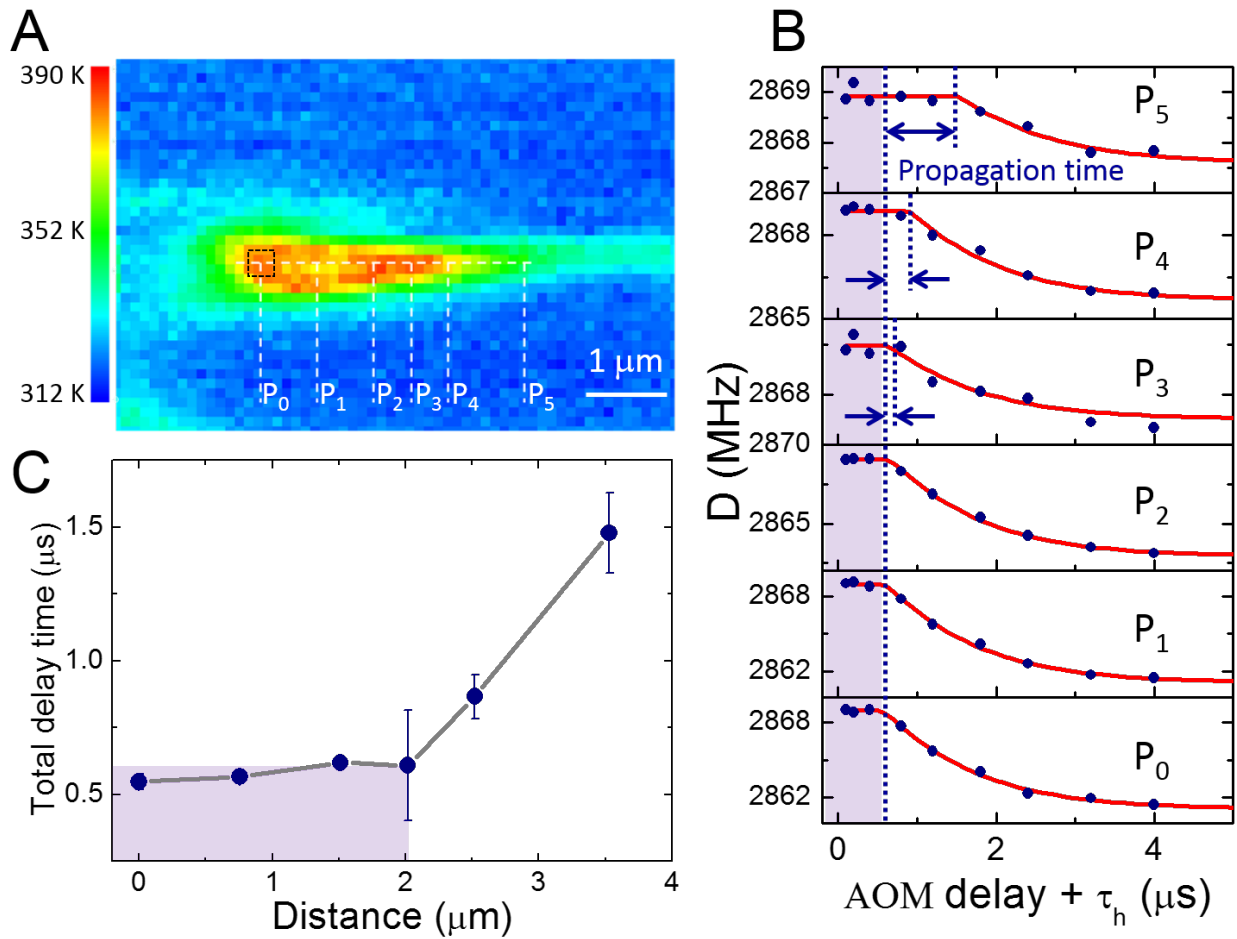

**Figure S5 | Mechanism of NIR heating.** (**A**) Two-dimensional heating image. Temperature measured by an ND (indicated by the black box) as a function of the position of the NIR laser beam. To speed up the imaging process, three-point CW ODMR was used. (**B**) Zero-field splitting $D$, extracted from full ODMR spectra, as a function of the NIR heating pulse duration for 5 different NIR heating positions (marked in **A**). Pulse sequence in Fig. 1C of the main text was used, with waiting time between the heating pulse and microwave pulse $t_w = 0$ μs and cooling time $t_c = 2$ μs before the readout. The shadowed region indicates the AOM delay. (**C**) The delay time for the ND temperature to rise after the NIR pulse (deduced from the fitting curve in **B**) as a function of the distance between the ND and the NIR laser focus spot. The 500-ns minimum value was due to the AOM delay.

## 5. Mechanism of HiT-ODMR

Figure S6 presents the ODMR spectra of an ND with the temperatures independently set during spin polarization and readout. The pulse sequences are presented in Fig. S6A. For the NIR power used, the stationary temperature was above 700 K ($D < 2815$ MHz). At such a high temperature, both the fluorescence of NV centers and the contrast were suppressed, and there was no ODMR signal when both the polarization and readout were carried out at high temperature (Fig. S6B, line 1). The fluorescence signal could be recovered when the readout was carried out at $T < 550$ K, but there was still no ODMR signal if the polarization pulse was applied at high temperature (line 3). The ODMR signal was barely observed at 2806 MHz when the spin was polarized at low temperature but read out at high temperature (line 2). The observed ODMR signal indicates that the spin polarization was well preserved during the fast heating and cooling processes. Optimal photon counts and ODMR contrast were achieved by performing both polarization and readout at low temperature (Fig. S6B, line 4). Thus, such a sequence (Fig. 1C) was used in the experiments shown in the main text.

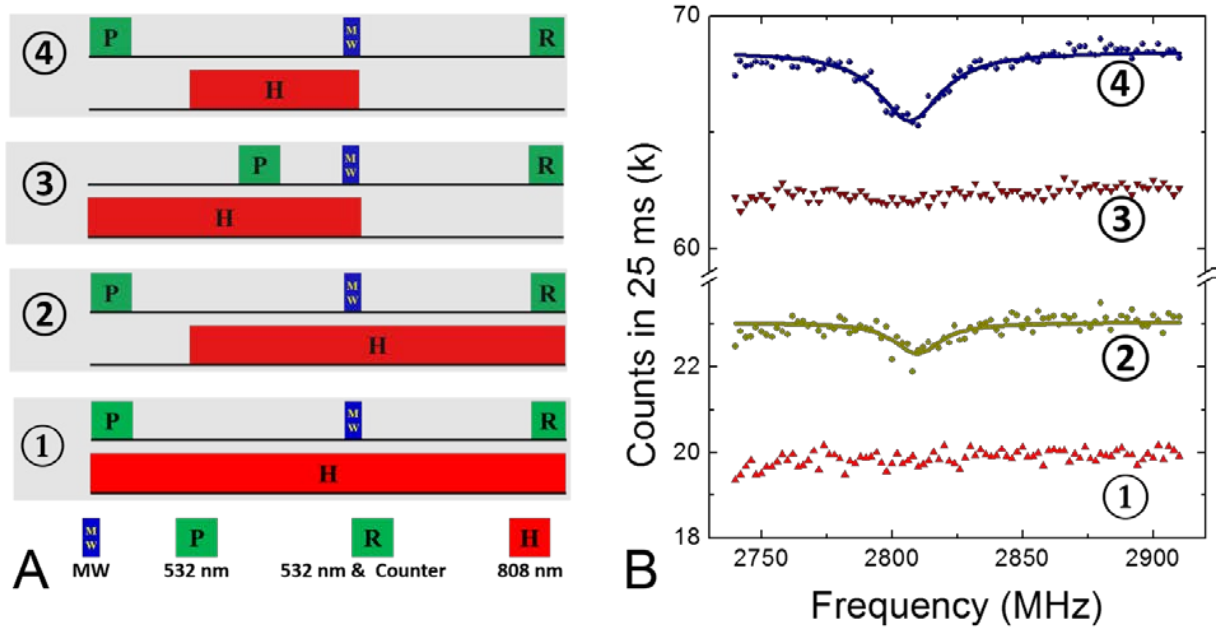

**Figure S6 | Mechanism of HiT-ODMR.** (**A**) Pulse sequences. The NV center spins were polarized (P) and read out (R) by the green laser and manipulated with the microwave pulse (MW). The ND was heated by an NIR pulse (H) with tunable duration and delay time. Sequence 1: polarization at high $T$, readout at high $T$; Sequence 2: polarization at low $T$, readout at high $T$; Sequence 3: polarization at high $T$, readout at low $T$; Sequence 4: polarization at low $T$, readout at low $T$. The MW pulse was always applied during the heating pulse (when the temperature was the highest). (**B**) ODMR spectra measured with the pulse sequences illustrated in **A**. The same integration time (25 ms for each MW frequency) was used for the 4 spectra, and the difference in counts was caused both by different emission efficiencies at different temperatures (1 vs 3 and 2 vs 4) and by different spin polarization rates at different temperatures (1 vs 2 and 3 vs 4).

## 6. Calibration of temperature at zero-field

The temperature below 700 K was calibrated with the *D-T* relation from literature (*1*), $D(T) = a_0 + a_1 T + a_2 T^2 + a_3 T^3$ with $a_0 = (2.8697 \pm 0.0009)$ MHz, $a_1 = (9.7 \pm 0.6) \times 10^{-2}$ GHz/K, $a_2 = (-3.7 \pm 0.1) \times 10^{-4}$ MHz/K$^2$, and $a_3 = (1.7 \pm 0.1) \times 10^{-7}$ MHz/K$^3$.

For temperature higher than 700 K, we measured the ZFS *D* at different waiting times $t_w$ (see Fig. 1C of main text and Fig. S7 for the pulse sequence) and then used extrapolation to determine the temperature right after the heating pulse with the assumption of exponential cooling (verified in Fig. S8).

The ODMR spectra at different $t_w$ were measured in a cycling manner (sequence in Fig. S7), so as to minimize possible effects due to aging of carbon films (which would reduce the heating efficiency), drifting of NIR laser focus spot, and laser power fluctuations. The protocol is as follows.

---

{

[P/R→H→wait($t_w$=0)→$\omega_1$→P/R→H→wait($t_w$=$t_1$)→$\omega_1$→… P/R→H→wait($t_w$=$t_n$)→$\omega_1$→P/R] x *M*,

[P/R→H→wait($t_w$=0)→$\omega_2$→P/R→H→wait($t_w$=$t_1$)→$\omega_2$→… P/R→H→wait($t_w$=$t_n$)→$\omega_2$→P/R] x *M*,

… …

[P/R→H→wait($t_w$=0)→$\omega_K$→P/R→H→wait($t_w$=$t_1$)→$\omega_K$→… P/R→H→wait($t_w$=$t_n$)→$\omega_K$→P/R] x *M*,

} x *N*

P/R – 532 nm laser pulse for spin polarization and readout

H – heating laser pulse (power fixed in the whole unit {…} x *N*),

wait($t_w$=$t_i$) – a waiting period of $t_i$ (*i* =1,2, …, *n*) between the heating pulse and the MW pulse

$\omega_k$ – MW pulse with frequency $\omega_k$ (fixed within each unit of […] x *M*)

*M* – repetition times of the unit […]

*N* – repetition times of the unit {…}.

---

We chose *n* = 5 and $t_w$ = {-0.2, 0.3, 0.6, 1.0, 1.4, 2.2} μs (excluding the NIR AOM delay) for experiments of lower or close to 700 K. For experiments with much higher temperatures, a longer delay was used, e.g., *n* = 5 and $t_w$ = {-0.2, 0.6, 1.0, 1.4, 2.2 2.6} μs (excluding the NIR AOM delay). Each unit of measurement for the 6 $t_w$'s lasted for about 100 μs. For each MW frequency the measurement lasted for about 20 ms (i.e., *M* = 200).

We have verified the method by comparing the temperature obtained by extrapolation with that determined by the *D-T* relation in Ref. (*1*) for *T* < 700 K (see Fig. 1E of main text). The examples of temperature calibration are shown in Fig. S8 and the results are summarized in Table S1. All the cooling curves are well fit with exponential decay functions with nearly the same cooling time except for the case of OD02. The differences between the cooling rate for OD02 and the others might be caused by the graphitization of the amorphous carbon film at high temperature. The highest temperature recorded was 1004 ± 24 K.

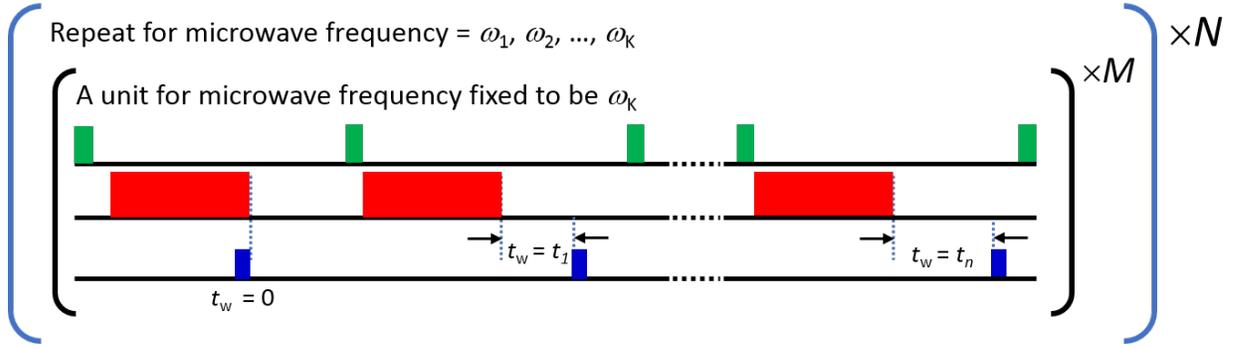

**Figure S7 | Pulse sequence for temperature calibration.** For each MW frequency, the following sequence was cycled $M$ times: First, the ODMR of the highest temperature was measured (MW pulse applied right at the end of the NIR heating pulse, $t_w = 0$); then ODMR were measured with the MW pulse applied with certain delay intervals ($t_w = t_1, t_2, \ldots, t_n$). The total time to run the sequence for $n = 5$ was about 100 µs, which was much shorter than the frequency sweep time of the microwave source (SMIQ03B, 20 ms was used for each frequency).

**Table S1 | Temperature calibration for an ND (NDa2) corresponding to ODMR data in Fig. S8.** For each NIR laser power (controlled by the OD-number), the zero-field splitting $D$ measured right at the end of the heating pulse ($t_w = -0.2$ µs) was converted to temperature $T_{0,\,DT}$ with the $D$-$T$ relation in Ref. (*1*) if it is < 700 K. The temperature right at the end of the heating pulse obtained by extrapolation of the exponential cooling curve in Fig. S8B is denoted as $T_{0,\,E}$. The temperatures obtained by the two methods agree well with each other.

|  | **OD11** | **OD08** | **OD05** | **OD03** | **OD02** |
|---|---|---|---|---|---|
| **Cooling time (µs)** | $1.2 \pm 0.1$ | $1.3 \pm 0.1$ | $1.47 \pm 0.03$ | $1.3 \pm 0.01$ | $0.79 \pm 0.02$ |
| **$D$ (MHz)** | $2859.1 \pm 0.4$ | $2842.0 \pm 0.4$ | $2824.9 \pm 0.3$ | $2801.5 \pm 0.6$ | $2758 \pm 1$ |
| **$T_{0,\,DT}$ (K)** | $409 \pm 5$ | $533 \pm 5$ | $639 \pm 4$ | N.A. | N.A. |
| **$T_{0,\,E}$ (K)** | $409 \pm 18$ | $536 \pm 26$ | $648 \pm 15$ | $771 \pm 4$ | $1004 \pm 24$ |

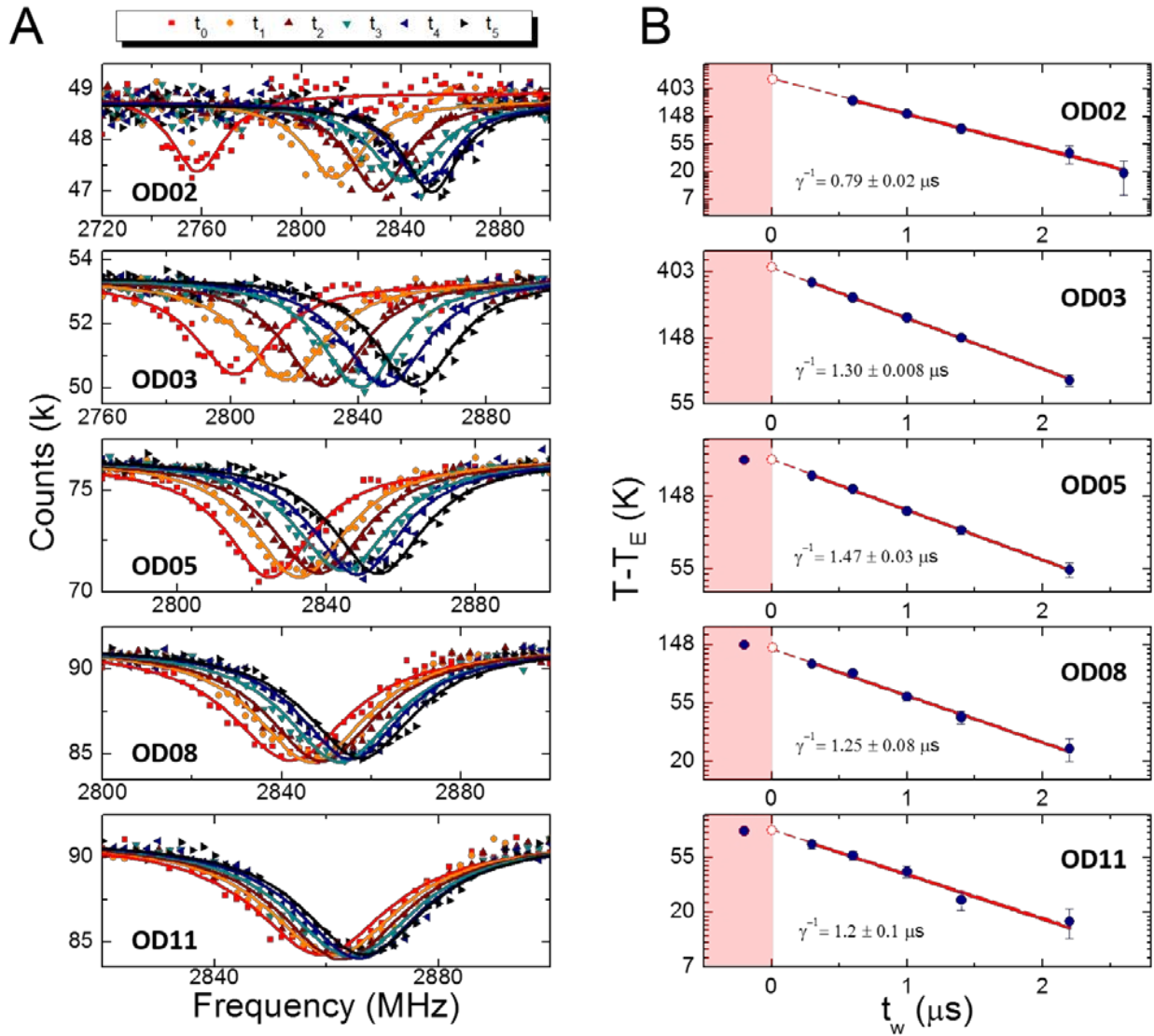

**Figure S8 | Temperature calibration by extrapolation of exponential cooling.** (**A**) ODMR spectra for various waiting times (indicated by different symbols), $t_w$ = {-0.2, 0.6, 1.0, 1.4, 2.2, 2.6} μs for OD02, and $t_w$ = {-0.2, 0.3, 0.6, 1.0, 1.4, 2.2} μs for the others. The pulse sequence in Fig. S7 was used. No external magnetic field was applied. The NIR laser power was increased from bottom to top, which was tuned by an optical attenuator (OD0 means full power). The ODMR spectra were fitted with a Lorentzian dip to extract the ZFS *D*. (**B**) The temperature of the ND (relative to the environment temperature $T_E$) as a function of the waiting duration $t_w$ for various NIR laser powers. The temperature lower than 700 K (solid symbols) was obtained using the *D-T* relation in Ref. (*1*). All the data for $t_w > 0$ are well fit with an exponential decay function. By extrapolation, the temperature immediately after the heating pulse was obtained (red open circles). The plateau before the end of the heating pulse (marked by the vertical dotted line) was due to the AOM delay (≈200 ns).

## 7. Spin coherence at room temperature

The spin coherence of ensemble NV centers in NDs at room temperature is shown in Fig. S9. An external magnetic field was applied to lift the degeneracy of the 4 crystallographic NV orientations. Typical spin relaxation time ($T_1$) was a few hundreds of microseconds. The spin coherence time, $T_2 = 0.91 \pm 0.05$ μs measured by spin echo and $T_2^* = 65 \pm 2$ ns measured by free-induction decay (FID), were much shorter than those of ensemble NV centers in high-purity bulk diamond (*3*).

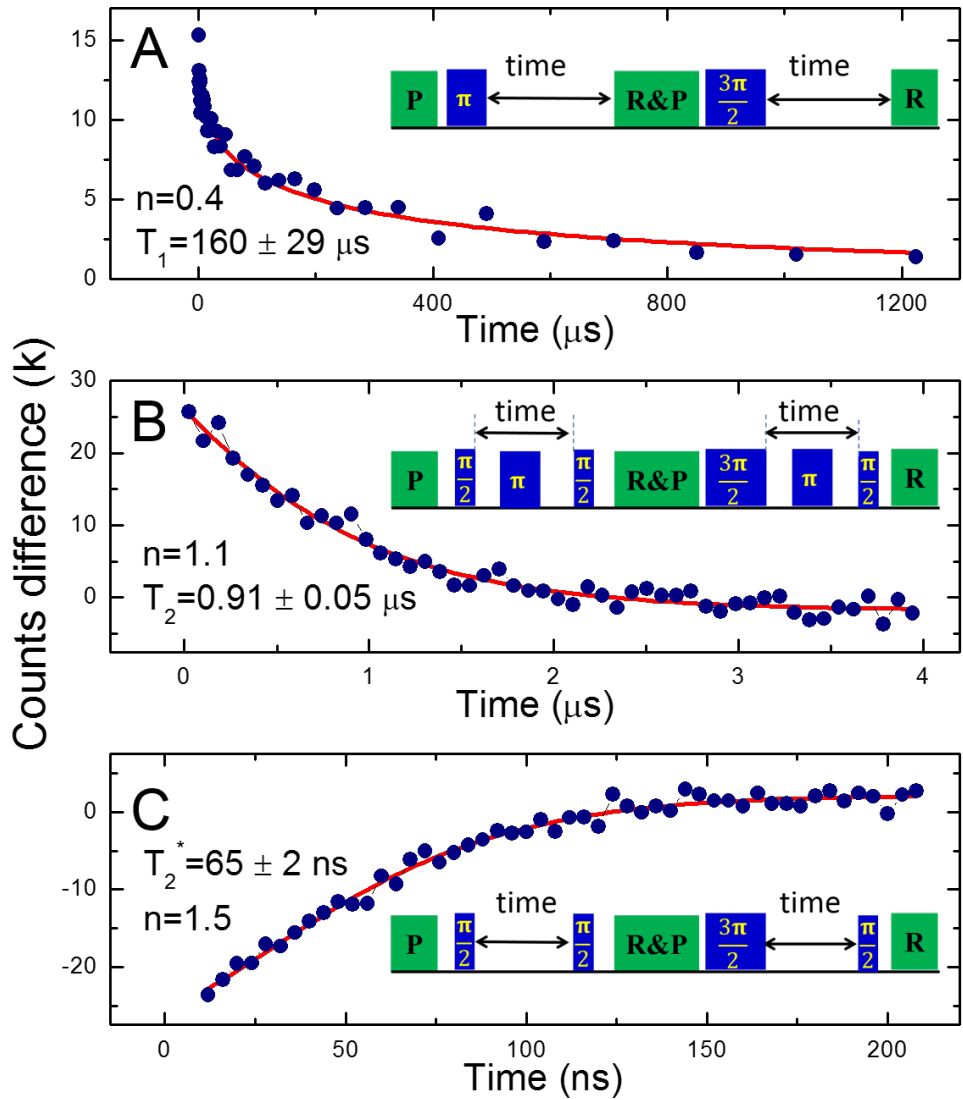

**Figure S9 | Room temperature spin coherence of ensemble NV centers in an ND.** (**A**) Spin relaxation (**B**) Spin echo and (**C**) Free-induction decay (FID) signals of the ensemble NV centers. An external magnetic field of about 80 Gauss was applied to lift the degeneracy of the NV centers of 4 different crystallographic orientations. Insets: pulse sequences for each measurement. To eliminate count fluctuations that are not related to spin process (e.g. laser power fluctuation and charge dynamics), the difference between the photon counts for the $m_s=0$ and $m_s = -1$ initial states was taken as the signals.

## 8. Temperature calibration in spin coherence measurement

To estimate the ND temperature in spin coherence measurements (data shown in Fig. 2A and 2D of the main text), we measured the full ODMR spectrum (Fig. S10A) under the same NIR laser power and duty ratio. Figure S10B shows the resonant frequencies (peak 1 and peak 4 in Fig. S10A) of NV centers in one orientation. The zero-field splitting $D$ was calculated by averaging the two resonant peaks (the effect of the transverse field (which was less than 10 Gauss), if any, would be much weaker than the variation of $D$). And the calibrated $D$-$T$ relation (Fig. 1E of the main text) was used to estimate the ND temperature.

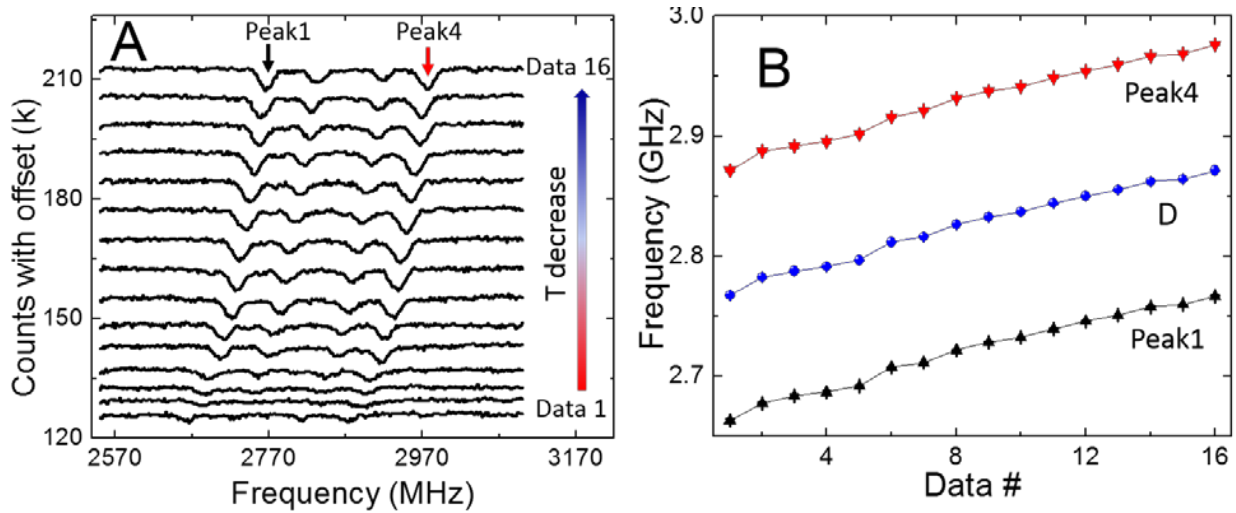

**Figure S10 | HiT-ODMR spectra for coherence measurement.** (**A**) ODMR spectra of NDa2 at different temperature. An external magnetic field of 38 Gauss was applied. The resonance peak 1 was used in Rabi oscillation and spin relaxation measurements (data in Fig. 2A and 2D of the main text). The ND temperature was calibrated with the $1^{st}$ and the $4^{th}$ resonances, which were associated with the NV centers with the smallest angle (among the 4 crystallographic orientations) from the external magnetic field. (**B**) The frequencies of peak 1 and 4 in **A** and their average (i.e., $D$) for Data # from 1 to 16 (corresponding to curves from bottom to top in **A**).

## 9. Measurement of magnetization of single Ni NPs.

The measurement protocol is as follows.

---

For each cooling process (temperature $T_\text{I} > T_\text{II} > \ldots > T_M$):

$\{[\text{P/R} \rightarrow T_\text{I} \rightarrow \omega_1] \times M' \rightarrow [\text{P/R} \rightarrow T_\text{I} \rightarrow \omega_2] \times M' \rightarrow \ldots \rightarrow [\text{P/R} \rightarrow T_\text{I} \rightarrow \omega_K] \times M'\} \times N,$

$\{[\text{P/R} \rightarrow T_\text{II} \rightarrow \omega_1] \times M' \rightarrow [\text{P/R} \rightarrow T_\text{II} \rightarrow \omega_2] \times M' \rightarrow \ldots \rightarrow [\text{P/R} \rightarrow T_\text{II} \rightarrow \omega_K] \times M'\} \times N,$

… …

$\{[\text{P/R} \rightarrow T_M \rightarrow \omega_1] \times M' \rightarrow [\text{P/R} \rightarrow T_M \rightarrow \omega_2] \times M' \rightarrow \ldots \rightarrow [\text{P/R} \rightarrow T_M \rightarrow \omega_K] \times M'\} \times N$

P/R – 532-nm laser pulse for spin polarization and readout

$T_m$ ($m$ = I, II, …, M) – temperature when the MW pulse was applied (controlled by the waiting duration $t_w = t_m$ after an NIR laser pulse). $T_m$ is fixed in each unit of $\{\ldots\} \times N$

$\omega_k$ ($k$ = 1, 2, …, K) – MW pulse with frequency $\omega_k$

$M'$ – repetition times of each unit $[\ldots]$

---

The magnetic field from the Ni NP induced splitting and broadening (due to the field gradient within the ND) (see Fig. 3A of main text). We used the splitting of the ODMR spectra to indicate the magnetic field from the nearby NiNP. As summarized in Fig. S12, the Ni NP was demagnetized at temperature higher than 615 ±4 K, which agrees well with the Curie temperature $T_C$ of bulk nickel, 627 K.

The spontaneous magnetization of this single Ni NP was tracked for 10 cooling processes (Round 1 to 10 in Fig. S12, and Round 1 to 4 also shown in Fig. 3B of main text). The magnetization was nearly the same if the Ni NP was kept at temperature lower than $T_C$, and new magnetic states could spontaneously emerge if the Ni NP had been heated to a temperature higher than $T_C$ (completely demagnetized, $T > 615$ K). Round 7 was measured with the NIR laser turned off and the ODMR spectra recorded repetitively, in order to estimate the statistic error of this experiment. The difference in ODMR splitting (e.g., that between Round 3 and 4) was much larger than the statistic error.

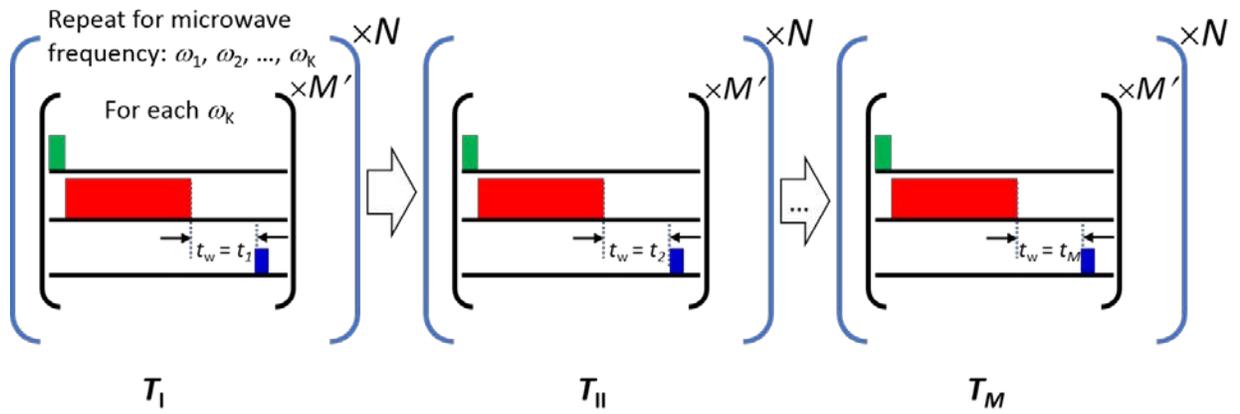

**Figure S11 | Pulse sequence for the Ni NP experiments.** The magnetization of a Ni NP was measured in a round of cooling process. The highest temperature was set by the power of the NIR laser pulse (saturated heating). The different temperatures ($T_I > T_{II} > \ldots > T_M$) were controlled by the waiting duration $t_w$. For each temperature $T_m$, the zero-field ODMR spectrum was measured in a frequency-cycling manner.

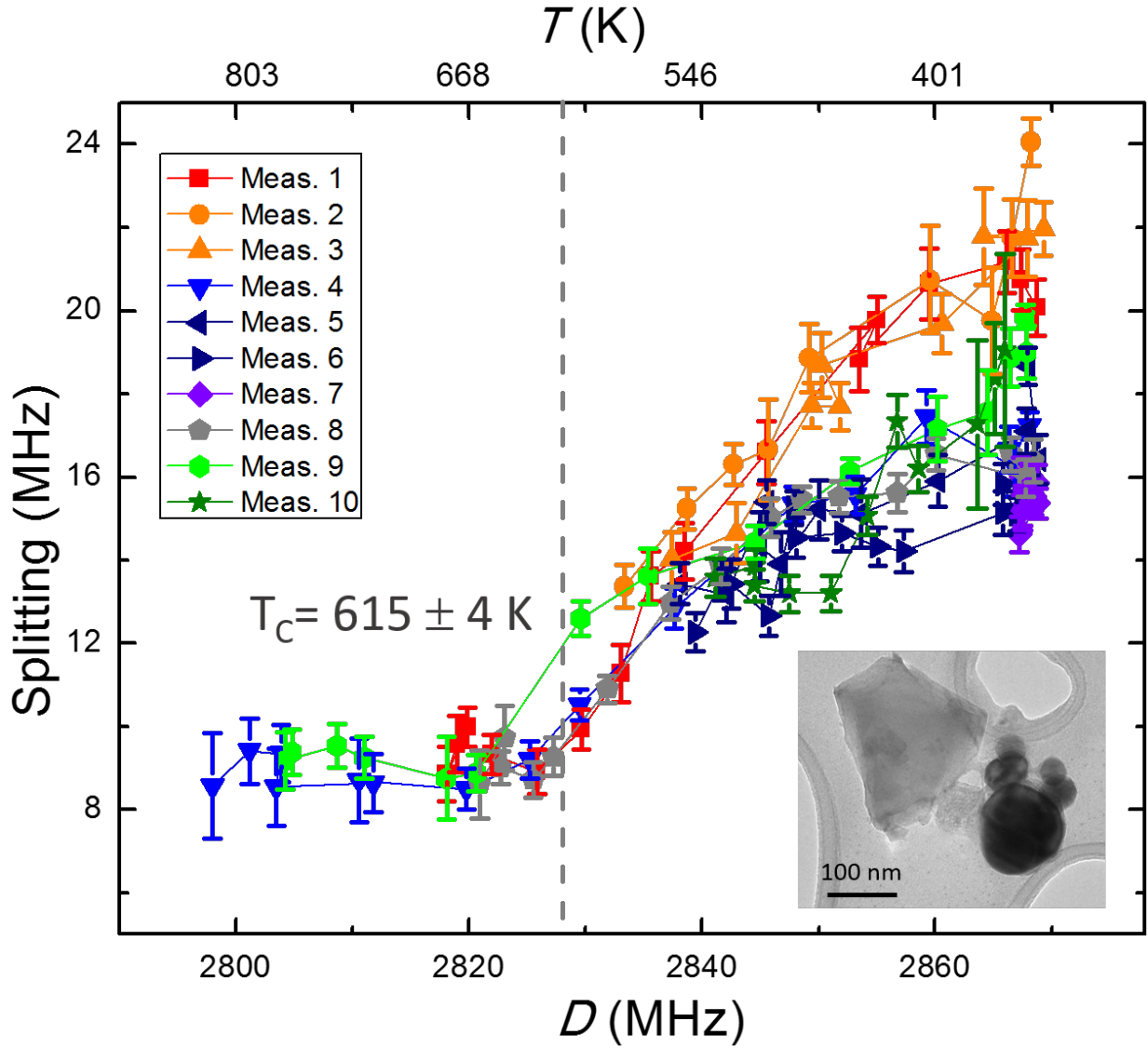

**Figure S12 | Spontaneous magnetization of a single Ni NP.** The ODMR splitting induced by the Ni NP was measured as a function of the ZFS $D$ (corresponding temperature shown in the top axis). The measurement was carried out from round 1 to round 10 (indicated by different symbols), each of which contained a sequence shown in Fig. S11. Round 1, 4, 8, and 9 were started at a temperature higher than $T_c$ (about 615 K, corresponding to $D = 2829$ MHz, indicated by the vertical dotted line). Round 7 was repetitive measurement at room temperature, which gives the statistic error of this method. Error bars are fitting error of the ODMR spectra (fitting by a double-peak Lorentzian function, see Fig. 3A of the main text). Inset: TEM image of the Ni NP and the ND on the carbon film that were measured.

## 10. Equations for sensitivities

The magnetic field sensitivity ($\eta_{B,cw}$) of an ND with ensemble NV centers was estimated by the following formula (4)

$$\eta_{B,cw} = \frac{h}{g\mu_B} \frac{\Delta\omega}{C\sqrt{L}}, \tag{S3}$$

where $L$ is the photon count rate, $h$ is the Planck constant, $g = 2.0$ is the Lande g-factor, $\mu_B$ is the Bohr magneton, and $\Delta\omega$ is the width of the ODMR resonant dip. In the Ni NP experiment, the experimental measured values were: $L = 8$ Mps, $\Delta\omega = 10$ MHz, and $C = 0.05$. And the estimated $\eta_{B,cw} = 2.5$ µT Hz$^{-1/2}$.

The temperature sensitivity ($\eta_{T,cw}$) of a ND nanothermometer was estimated by the following formula (2)

$$\eta_{T,cw} = \frac{\Delta\omega}{C\sqrt{L}|dD/dT|}. \tag{S4}$$

For $dD/dT = 240$ KHz/K at temperature close to 1000 K, the sensitivity $\eta_{T,cw} = 250$ mk Hz$^{-1/2}$.

## SUPPLEMENTARY REFERENCES

1. D. M. Toyli *et al.*, Measurement and control of single nitrogen-vacancy center spins above 600 K. *Phys. Rev. X*. **2** (2012).
2. N. Wang *et al.*, Magnetic Criticality Enhanced Hybrid Nanodiamond Thermometer under Ambient Conditions. *Phys. Rev. X*. **8**, 011042 (2018).
3. P. L. Stanwix *et al.*, Coherence of nitrogen-vacancy electronic spin ensembles in diamond. *Phys. Rev. B*. **82**, 201201 (2010).
4. L. Rondin *et al.*, Magnetometry with nitrogen-vacancy defects in diamond. *Reports Prog. Phys.* **77** (2014).